
\magnification=1200
\baselineskip=14pt plus 1pt minus 1pt
\tolerance=1000\hfuzz=1pt


\font\bigfont=cmr10 scaled\magstep3
\font\secfont=cmr10 scaled\magstep1

\def\section#1#2{\vskip32pt plus4pt \goodbreak \noindent{\bf\secfont#1. #2}
	\xdef\currentsec{#1} \global\eqnum=0 \global\thmnum=0}
\def\appendix#1#2{\vskip32pt plus4pt \goodbreak \noindent
		{\bf\secfont{Appendix }#2}
	\xdef\currentsec{#1} \global\eqnum=0 \global\thmnum=0}

\def\leftheader{}
\def\rightheader{}
\def\header{\headline={\tenrm\ifnum\pageno>1
                        \ifodd\pageno\rightheader
                         \else\noindent\leftheader
			  \fi\else\hfil\fi}}

\newcount\thmnum
\global\thmnum=0
\def\prop#1#2{\global\advance\thmnum by 1
	\xdef#1{Proposition \currentsec.\the\thmnum}
	\bigbreak\noindent{\bf Proposition \currentsec.\the\thmnum.}
	{\it#2} }
\def\defin#1#2{\global\advance\thmnum by 1
	\xdef#1{Definition \currentsec.\the\thmnum}
	\bigbreak\noindent{\bf Definition \currentsec.\the\thmnum.}
	{#2} }
\def\lemma#1#2{\global\advance\thmnum by 1
	\xdef#1{Lemma \currentsec.\the\thmnum}
	\bigbreak\noindent{\bf Lemma \currentsec.\the\thmnum.}
	{\it#2} }
\def\thm#1#2{\global\advance\thmnum by 1
	\xdef#1{Theorem \currentsec.\the\thmnum}
	\bigbreak\noindent{\bf Theorem \currentsec.\the\thmnum.}
	{\it#2} }
\def\cor#1#2{\global\advance\thmnum by 1
	\xdef#1{Corollary \currentsec.\the\thmnum}
	\bigbreak\noindent{\bf Corollary \currentsec.\the\thmnum.}
	{\it#2} }
\def\conj#1#2{\global\advance\thmnum by 1
	\xdef#1{Conjecture \currentsec.\the\thmnum}
	\bigbreak\noindent{\bf Conjecture \currentsec.\the\thmnum.}
	{\it#2} }

\def\proof{\medskip\noindent{\it Proof. }}

\newcount\eqnum
\global\eqnum=0
\def\num{\global\advance\eqnum by 1
	\eqno({\rm\currentsec}.\the\eqnum)}
\def\eqalignnum{\global\advance\eqnum by 1
	({\rm\currentsec}.\the\eqnum)}
\def\ref#1{\num  \xdef#1{(\currentsec.\the\eqnum)}}
\def\eqalignref#1{\eqalignnum  \xdef#1{(\currentsec.\the\eqnum)}}

\def\title#1{\centerline{\bf\bigfont#1}}

\newcount\subnum
\def\Alph#1{\ifcase#1\or A\or B\or C\or D\or E\or F\or G\or H\fi}
\def\subsec{\global\advance\subnum by 1
	\vskip12pt plus4pt \goodbreak \noindent
	{\bf \currentsec.\Alph\subnum.}  }
\def\newsubsec{\global\subnum=1 \vskip6pt\noindent
	{\bf \currentsec.\Alph\subnum.}  }
\def\today{\ifcase\month\or January\or February\or March\or
	April\or May\or June\or July\or August\or September\or
	October\or November\or December\fi\space\number\day,
	\number\year}
\input amssym.def
\input amssym.tex

\def\ol{\overline}
\def\ul{\underline}

\def\tr{\mathop{\rm tr}\nolimits}

\def\supp{\mathop{\rm supp}\nolimits}

\def\bC{{\Bbb C}}

\def\bZ{{\Bbb Z}}
\def\bfA{{\bf A}}

\def\cA{{\cal A}}

\def\cC{{\cal C}}
\def\cD{{\cal D}}
\def\cF{{\cal F}}
\def\cH{{\cal H}}
\def\cI{{\cal I}}
\def\cL{{\cal L}}

\def\cS{{\cal S}}

\def\cX{{\cal X}}
\def\fA{{\frak A}}

\def\fH{{\frak H}}

\chardef\o="1C

\header
\def\leftheader{\centerline{C. KING and A. LESNIEWSKI}}
\def\rightheader{\centerline{QUANTUM CODING}}

\def\intsec{I}
\def\quantsourcesec{II}
\def\ergodicsec{III}
\def\quantentsec{IV}
\def\equipartsecsec{V}
\def\markovapp{A}

\def\csrc{(\cX^\infty,T,\mu)}
\def\qsrc{(\fA,\alpha,\tau)}
\def\la{\langle}
\def\ra{\rangle}
\def\supp{{\rm supp}}

{\baselineskip=12pt
\nopagenumbers
\line{\hfill \bf HUTMP 95/441}
\line{\hfill \today}
\vfill
\title{Quantum sources and a quantum coding theorem}
\vskip 1in
\centerline{{\bf Christopher King}$^*$
and {\bf Andrzej Le\'sniewski}$^{**}$\footnote{$^1$}
{Supported in part by the National Science Foundation under grant
DMS-9424344}}
\vskip 12pt
\centerline{$^*$Department of Mathematics}
\centerline{Northeastern University}
\centerline{Boston, MA 02115, USA}
\centerline{\tt king@neu.edu}
\vskip 12pt
\centerline{$^{**}$Lyman Laboratory of Physics}
\centerline{Harvard University}
\centerline{Cambridge, MA 02138, USA}
\centerline{\tt lesniewski@huhepl.harvard.edu}
\vskip 1in\noindent
{\bf Abstract}.
We define a large class of quantum sources and prove a quantum analog
of the asymptotic equipartition property. Our proof relies
on using local measurements on the quantum source to
obtain an associated classical source. The classical source provides
an upper bound for the dimension of the relevant subspace of the
quantum source, via the Shannon-McMillan noiseless coding theorem.
Along the way we derive a bound for the von Neumann entropy
of the quantum source in terms of the Shannon entropy of the classical
source, and we provide a definition of ergodicity of the quantum
source. Several explicit models of quantum sources are also presented.
\vfill\eject}

\section\intsec{Introduction}

\newsubsec
The possibility of building a quantum computer
has stimulated new interest in the quantum analog of
classical information theory (see [Be] for an introduction
and review of current ideas). Shannon, McMillan, Khinchin
and others provided a firm foundation for the classical
theory, and used the mathematics of stochastic
processes to prove important theorems. In particular they
obtained limits on the amount
of information that can be transmitted through a
channel. Although a quantum computer does not yet exist,
it is reasonable to suppose that similar issues of channel
capacity are relevant to its operation. So it is interesting
to investigate this question, using our current
understanding of quantum mechanics.

There has been much work done on this and related questions.
In particular,
Schumacher stated and proved a capacity result for a
quantum channel in [S], [JS]. Our particular interest is
in the {\it extended} quantum signal source described by Schumacher.
This is a quantum system whose state space is a (tensor) product
of many copies of one fundamental state space $M$.
The source produces a signal which is encoded by a state in $M$;
the ensemble of possible signals is represented by a density
operator $\rho$ on $M$. The extended
source corresponds to a sequence of such states, and has a
natural interpretation as a message. The probabilistic character
of the message is contained in the density operator on the
tensor product of copies of $M$. One choice of density operator is
$\rho \otimes \dots \otimes \rho$. This corresponds to independent
signals at all times, and there are no correlations between
signals in the message. For this reason we call this a quantum
Bernoulli source (a precise definition is provided in section II).
Schumacher proves his quantum noiseless coding theorem for such
a source. He introduces the notion of fidelity of a quantum
channel, and states his results in terms of this. On the most basic
level his results show that the state space for the extended source
has a relevant subspace
whose dimension is determined by the von Neumann entropy of
$\rho$. As far as the information content of the source is concerned,
the rest of the state space can be ignored.

\subsec
Our main interest in this paper is to extend Schumacher's result
to a large class
of quantum sources. We define precisely what we mean by this
in section II. Roughly speaking, we consider sources which
allow correlations on all time scales between signals in a message.
In classical information theory the coresponding result is the
Shannon-McMillan theorem, which shows that under very general
assumptions the ensemble of all possible messages can be split
into a relevant and an irrelevant part. The criterion for splitting
is provided by the entropy of the source. The corresponding result
for the quantum theory should be a splitting of the state space
into a relevant subspace and an irrelevant subspace, with the
von Neumann entropy as the criterion. This is precisely what
Schumacher proved for the quantum Bernoulli source.
We obtain such a splitting in the general case, and derive
an estimate for the dimension of the relevant subspace,
by computing the entropy of a classical source which is obtained by
making measurements on the quantum system.

To summarize, we show that a general quantum source can be encoded
by a quantum system whose dimension is smaller than the original.
We give estimates for the dimension of the reduced space, based on
the results of local measurements of the source. For the case
of a quantum source emitting orthogonal states
our estimate is tight, and for a Bernoulli source
reproduces Schumacher's result.
We also derive an inequality relating this experimental entropy
to the true von Neumann entropy.

\subsec
The paper is organized as follows. In section II we define quantum
sources, and recall some standard results about the construction
of infinite quantum systems. We also provide some explicit
examples which serve to illustrate our ideas. In section III we
recall the classical notion of an ergodic source, and propose a
definition of ergodicity for a quantum source. As a check we prove
that a quantum Bernoulli source is ergodic. In section IV we recall
the notion of quantum entropy. In section V we pursue the quantum
analog of the Shannon-McMillan theorem. To do this we use the
notion of a positive operator valued measure, and
construct an associated classical source. The entropy of this
associated source satisfies a lower bound involving the von
Neumann entropy of the quantum source. It also determines the
dimension of a relevant subspace of the quantum system which
in turn yields our quantum noiseless coding theorem.

\section\quantsourcesec{Quantum sources}

\newsubsec
According to the mathematical theory of information, see e.g.
[A], [B], [K], a (classical) {\it source} is a stochastic process. In this
paper, we are concerned with discrete sources which can be described
as follows. We are given a finite set $\cX$, called the alphabet, and
consider the space $\cX^\infty$ of all infinite sequences $\ul x=
\{x_n\}_{n\in\bZ}$, called messages. The time evolution is given
by the shift $T:\;\cX^\infty\rightarrow\cX^\infty$ defined by
$(T\ul x)_n:=x_{n+1}$. The stochastic character of the process is
governed by a probability measure $\mu$ on $\cX^\infty$. The triple
$\csrc$ is called a source. We say that the source is
{\it stationary}, if $T$ preserves the measure $\mu$.

We begin by describing informally the properties of a
quantum source. In the next section we will provide a rigorous
mathematical description. By analogy with the classical case,
a quantum source will be a triple consisting of {\it quantum messages},
the {\it time shift}, and a {\it probability distribution} for the
messages. For technical reasons, it is useful to describe the
space of quantum messages (which is a linear space) by the algebra
of observables on it.

A {\it quantum source} sends a series of signals, each of which is a
vector in a finite dimensional Hilbert space $\cH$. We assume that the
source is discrete, i.e. each signal is an element of a finite
set $\cS=\{|\psi_1\ra,\ldots, |\psi_s\ra\}$ of normalized vectors in $\cH$.
To avoid unnecessary redundancy, we assume that the space
$\cH$ is spanned by $\cS$, i.e. $\cH\simeq\bC^d$, where $d\leq s$.
We do not assume that the elements of $\cS$ are orthogonal to each
other or even linearly independent. Indeed, this is an important
difference from the classical situation which does not allow for
forming linear superpositions of states. Denote by $p_j$ the
{\it a priori} probability of the state $|\psi_j\ra$ being sent.
The density matrix corresponding to the ensemble of signals $\cS$
is then given by
$$
\rho=\sum_{1\leq j\leq s}p_j|\psi_j\ra\la\psi_j|.\ref{\rhodef}
$$
As a consequence of our assumptions, $\tr(\rho)=1$.
Clearly, an ensemble of signals $\cS$ and the associated {\it a priori}
probabilities determines uniquely the density matrix $\rho$. On
the other hand, any given density matrix corresponds to infinitely
many different sets of signals. For a discussion of this point,
see [HJW].

The observables associated with quantum signals are $d \times d$
hermitian matrices; more formally (we will need this viewpoint
shortly) they are
elements of the $\bC^*$-algebra $\cA=\cL(\cH)$ of linear operators
on $\cH$. Abusing slightly the language, we will refer to $\cA$ as
the algebra of observables. Using the language adopted in the operator
algebra approach to quantum physics, see e.g. [H], [BR], we define the
{\it state} on the algebra of observables $\cA$ associated with the
density matrix $\rho$ to be
$$
\tau_1(A):=\tr(A\rho) =
\sum_{1\leq j\leq s}p_j\la\psi_j|A|\psi_j\ra.\ref{\singlestate}
$$

\subsec
We now propose a formal definition of a quantum source. To this end,
we construct the state on the algebra of observables associated
with entire (infinite) quantum messages rather than individual signals.
This is technically somewhat delicate, as it involves infinite tensor
products of Hilbert spaces. Let $I\subset\bZ$ be a finite set of the
form $\{M,\;M+1,\ldots,N-1\;,N\}$, where $M<N$, i.e. $I$ is a finite
collection of consecutive integers. By $\cI$ we denote the partially
ordered set of all such $I$'s. We set $\fH_I:=\bigotimes_{j\in I}\cH_j$,
where $\cH_j=\cH$ for all $j\in I$, and define the corresponding
observable algebra $\fA_I:=\cL(\fH_I)\simeq\bigotimes^{|I|}\cL(\cH)$, where
$|I|$ denotes the number of elements in $I$. For $I\subset J$, there
is a natural embedding $\cL(\fH_I)\hookrightarrow\cL(\fH_J)$, and so we can
form the union $\fA_{loc}:=\bigcup_{I\in\cI}\fA_I$. The latter is a
normed algebra, and we refer to its elements as {\it local observables}.
Roughly, $\fA_{loc}$ is a collection of operators acting on the
infinite tensor product $\bigotimes_{j\in \bZ}\cH_j$;
every element of $\fA_{loc}$ acts as the identity on all but a  finite
number of factors in this product.
For a local observable $A\in\fA_{loc}$, we let $\supp(A)$ denote its
support, i.e the smallest $I\in\cI$ such that $A\in\fA_I$. The norm
closure $\fA$ of $\fA_{loc}$ is a $\bC^*$-algebra called the algebra
of {\it quasilocal observables}. These concepts are borrowed
from algebraic field theory and statistical mechanics, and we refer
the reader to [H] and [BR] for a thorough presentation.

We will use the net $\{\fA_I\}_{I\in\cI}$ of matrix algebras to
construct the Hilbert space of states of a quantum source. Assume
that we have a family $\{\Pi_I\}_{I\in\cI}$, $\Pi_I\in\fA_I$,
satisfying the following assumptions:

\item{$1^\circ.$} {Each $\Pi_I$ is a positive operator.}
\item{$2^\circ.$} {If $|I|=1$, then $\Pi_I=\rho$.}
\item{$3^\circ.$} {Let $I\subset J$ be such that $J\backslash I
\in\cI$. Then
$$
\tr_{\fH_I}\Pi_J=\Pi_{J\backslash I},\ref{\consistency}
$$
where $\tr_{\fH_I}$ denotes the partial trace over the factor
$\fH_I$ in the tensor product $\fH_J=\fH_I\otimes\fH_{J\backslash I}$
or $\fH_J=\fH_{J\backslash I}\otimes\fH_I$.}

\noindent
Note that the last condition implies, in particular, that $\tr\Pi_I=1$.
In other words, $\{\Pi_I\}_{I\in\cI}$ is a {\it consistent} family of
density matrices, and it can be thought of as a quantum mechanical
counterpart of a consistent family of cylinder measures.

For each $I\in\cI$ we define a state $\tau_I$ on $\fA_I$ by
$$
\tau_I(A):=\tr(A\Pi_I),\ref{\taui}
$$
and observe that $|\tau_I(A)|\leq \Vert A\Vert$, uniformly in $I$. The
consistency condition {\consistency} implies that $\tau_I$ is well
defined, and that the generalized
limit $\tau(A):=\lim_{I\nearrow\bZ}\;\tau_I(A)$ exists for all $A\in
\fA_{loc}$, and satisfies $|\tau(A)|\leq \Vert A\Vert$.
As a consequence,
$\tau$ can be uniquely extended to a state on the $\bC^*$-algebra
$\fA$ of quasilocal observables. We use the same symbol $\tau$ to
denote this extension.

Let $\fH,\pi$ be the GNS representation, see e.g. [BR], associated
with the state $\tau$. The Hilbert space $\fH$ is the state space
of the quantum source. If no confusion arises, we will write $A$
instead of $\pi(A)$.

The additive group $\bZ$ underlying the above construction plays the
role of (discrete) time translations. Its action on the algebra
of local observables is defined as follows:
$$
\fA_I\ni A\simeq A\otimes I\longrightarrow \alpha(A):=
I\otimes A\simeq A\in\fA_{I+1}.\ref{\alphadef}
$$
In other words, $\alpha$ pushes the observable to the right by
one unit of time. Clearly, $\Vert\alpha(A)\Vert=\Vert A\Vert$, and so
$\alpha$ has a unique extension to all of $\fA$ which we will
denote by the same symbol. The family of automorphisms $\{
\alpha^n\}_{n\in\bZ}$ defines then a representation of $\bZ$.
The triple $\qsrc$ is called a {\it quantum source}.

We say that the quantum source is {\it stationary}, if the state
$\tau$ is invariant under $\alpha$, i.e.
$$
\tau(\alpha(A))=\tau(A),\num
$$
for all $A\in\fA$. From now on we will be assuming that our source
is stationary. A standard result in operator algebras implies that
the automorphism $\alpha$ is unitarily implementable on the GNS
Hilbert space associated with the invariant state $\tau$, i.e.
$\alpha(A)=FAF^{-1}$ on $\fH$. We will call the unitary operator
$F$ a {\it quantum shift}.

Because of stationarity, we can always assume that $I\in\cI$ is of
the form $\{1,\ldots,n\}$. To simplify the notation, we will
write then $\Pi_n$ instead of $\Pi_I$, and $\tau_n$ instead of
$\tau_I$.

\subsec
The simplest example of a quantum source is a
{\it Bernoulli source}, which we now describe. As in the classical
case, a Bernoulli source produces messages which are sequences of
independent signals. Accordingly the family of density matrices
$\{\Pi_I\}$ is given as follows:
$$
\Pi_{I} = \otimes_{i \in I} {\rho}_{i}, \ref{\defBern}
$$
where each ${\rho}_{i}$ equals $\rho$. It immediately follows that the
consistency condition {\consistency} is satisfied, and that
the source is stationary. This is the class
of sources considered by Schumacher [S]. One special feature of a
Bernoulli source is that whenever ${\rm supp}(A) \cap {\rm supp}(B)
= \emptyset$, we have $\tau(AB) = \tau(A) \tau(B)$.

There are many examples of non-Bernoulli sources. We present here
a special class of stationary sources. These are all described by
a signal density matrix $\rho$ and another matrix
$R \in \cL(\cH \otimes \cH)$ satisfying
$$
\tr_{1}((\rho \otimes I)R) = \rho, \quad
\tr_{2}(R) = I.\ref{\Rprop}
$$
Here $I$ is the identity matrix on $\cH$, and we introduce the
notation $\tr_{i}$ for the
partial trace over the $i$th factor in the $n$-fold tensor product
$\otimes^{n}\cH$. The density matrices $\{\Pi_{n}\}$ are then
constructed recursively as follows:
$$
\eqalign{
{\Pi}_{1} &= \rho,\cr
{\Pi}_{n+1} &= {1 \over 2} ({\Pi}_{n} \otimes I_{1})
(I_{n-1} \otimes R) +
{1 \over 2} (I_{n-1} \otimes R) ({\Pi}_{n} \otimes I_{1}).\cr}
$$
\medskip\noindent
We have denoted by $I_{n}$ the identity matrix on the product
$\otimes^{n}\cH$. The consistency of this definition ({\consistency})
is immediate. The positivity of the density matrices $\Pi_{n}$ is a
further constraint on $R$. We have several explicit examples for which
the positivity can be proven. In the simplest situation the matrix $R$
satisfies the following additional conditions:
$$\eqalign{
R& \quad \geq \quad 0,\cr
[\rho \otimes I_{1}&, R] \quad = \quad 0,\cr
[I_{1} \otimes R, R& \otimes I_{1}] \quad = \quad 0.\cr}\ref{\commutingR}
$$

It follows readily that all the matrices $\Pi_{n}$ are positive, for all
$n \geq 1$. For example, suppose $\rho = \sum_{j=1}^{d} {\lambda}_{j}
P_{j}$, where $\{\lambda_{j}\}$ are the eigenvalues of $\rho$, and
$\{P_{j}\}$ are the corresponding orthogonal projections. Then we can
take $R = \sum_{j=1}^{d} P_{j} \otimes P_{j}$, and all the above properties
are easily seen to hold.

For our second example $\cH = {\bC}^{2}$, and we assume that $\rho$ is
strictly positive. Let $\{\sigma_{j}\}$ be the Pauli matrices:
$$
{\sigma}_{1} = \left(\matrix{0 & 1\cr 1 & 0\cr}\right), \quad
{\sigma}_{2} = \left(\matrix{0 & -i\cr i & 0\cr}\right), \quad
{\sigma}_{3} = \left(\matrix{1 & 0\cr 0 & -1\cr}\right).
$$
By choosing a suitable basis we can write
$$
\rho = {1 \over 2} I + {a \over 2} {\sigma}_{3},\ref{\twobytworho}
$$
where $|a| < 1$. Then we take the matrix $R$ to be
$$
R = I \otimes \rho + ({a \over 2}I - {1 \over 2}{\sigma}_{3})
\otimes (b{\sigma}_{1} + c{\sigma}_{2}).\ref{\twobytwoR}
$$
The consistency conditions are easily verified. Positivity of all matrices
$\{\Pi_{n}\}$ holds for $|b|,\,|c|$ sufficiently small; the proof is given
in the Appendix.

\subsec
The density matrix ${\Pi}_{n}$ can also be written in terms of the
states $\{{\psi}_{j}\}$ which span $\cS$ (compare {\rhodef}) as follows:
$$
{\Pi}_{n} = \sum_{1 \leq j_{1},\dots,j_{n} \leq s}
p_{j_{1},\dots,j_{n}} |{\psi}_{j_{1}}\rangle \langle {\psi}_{j_{1}}|
\otimes \ldots\otimes |{\psi}_{j_{n}}\rangle \langle {\psi}_{j_{n}}|.
$$
The numbers $p_{j_{1},\dots,j_{n}}$ satisfy obvious consistency
conditions. If the states $\{{\psi}_{j}\}$ are linearly independent,
then they are also non negative. Further, if the states are orthogonal,
the density matrices $\{{\Pi}_{n}\}$ commute. In this case, the
quantum source is equivalent to a classical source.

\section\ergodicsec{Ergodic quantum sources}

\newsubsec
According to the individual ergodic theorem [CFS], for any function
$f\in L^1(\cX^\infty)$, the sequence of time averages
$$
\la f\ra_N(\ul x):={1\over N}\sum_{0\leq n\leq N-1}
f(T^n\ul x)\ref{\ctimeav}
$$
converges almost everywhere to a limit $\ol f$. The function $\ol f$
is invariant under $T$. The source $\csrc$ is called {\it ergodic}, if
the only functions invariant under $T$ are constants. Consequently,
for an ergodic source $\ol f=\int_{\cX^\infty}f(\ul x)d\mu(\ul x)$.
This condition is one of the several equivalent statements which
could be used to define ergodicity, see e.g. [CFS]. There is no natural
concept of convergence almost everywhere in quantum mechanics.
Consequently we adopt a notion of convergence which is natural
for the algebra $\fA$. Our definition
of ergodicity of a quantum source is specifically
designed to suit the purposes of this paper, although it may have
broader applications.

We say that a quantum source is {\it ergodic} if the following condition
is satisfied. For any $A\in\fA$, the time averages
$$
\la A\ra_N:={1\over N}\;\sum_{0\leq n\leq N-1}\;
F^nAF^{-n}\ref{\qtimeav}
$$
converge in a suitable sense to the limit $\tau(A)I$, as $N\to\infty$.
Specifically, we require weak convergence, i.e. for all $\phi,\psi\in\fH$,
$$
\lim_{N\to\infty}\big(\phi,(\la A\ra_N-\tau(A)I)\psi\big)=0.
\ref{\timeavconvone}
$$
For $A\in\fA$, we let $[A]$ denote the corresponding element of
$\fH$. Substituting $\phi=[I]$ and $\psi=[B],\;B\in\fA_{loc}$ in
\timeavconvone , we find that for an ergodic source,
$$
\lim_{N\to\infty}\tau(\la A\ra_NB)=\tau(A)\tau(B),
\ref{\timeavconvtwo}
$$
for all $A,B\in\fA_{loc}$.

\subsec
The simplest example of an ergodic source is a Bernoulli source.

\prop\bernoulliergodicity{A quantum Bernoulli source is ergodic.}
\proof Let $A,B,C\in\fA_{loc}$. We assert that
$$
\tau(B^{\dag}\la A\ra_NC)\rightarrow\tau(B^{\dag} C)\tau(A),
\ref{\wconv}
$$
which is equivalent to \timeavconvone\ with $\phi=[B],\;\psi=[C]$.
To prove this, we observe that there is $n_0$ such that
$$
\tau(B^{\dag}F^nAF^{-n}C)=\tau(F^nAF^{-n})\tau(B^{\dag}C)=
\tau(A)\tau(B^{\dag}C),
$$
for $n>n_0$ (this follows from the fact that the supports of $A,B$,
and $C$ are all finite.) Consequently,
$$
\tau(B^{\dag}\la A\ra_N C)={1\over N}\;\sum_{0\leq n\leq n_0}\;
\tau(B^{\dag}F^nAF^{-n}C)+{{N-n_0-1}\over N}\;\tau(A)\tau(B^{\dag}C).
$$
But
$$
{1\over N}\;|\sum_{0\leq n\leq n_0}\;\tau(B^{\dag}F^nAF^{-n}C)|\leq
{{n_0+1}\over N}\;\Vert A\Vert\Vert B\Vert\Vert C\Vert\longrightarrow 0,
$$
and so $\lim_{N\to\infty}\tau(B^{\dag}\la A\ra_NC)=\tau(A)
\tau(B^{\dag}C)$.

The remainder of the proof is a series of straightforward
approximation arguments.

\noindent
$1^{\circ}$. We claim that \wconv\ holds for all $A,B,C\in\fA$.
To prove this, observe that for all $N$,
$$
\Vert\la A\ra_N\Vert\leq\Vert A\Vert.\ref{\timeavcont}
$$
For $A\in\fA$, let $A_j\in\fA_{loc}$ be such that $\Vert A-A_j\Vert
\rightarrow 0$, as $j\rightarrow\infty$. For any $B,C\in\fA_{loc}$, write
$$
\eqalign{
\tau\big(B^{\dag}\la A\ra_NC-\tau(A)B^{\dag}C\big)&=
\tau\big(B^{\dag}\la A_j\ra_NC-\tau(A_j)B^{\dag}C\big)\cr
&+\tau\big(B^{\dag}\la A-A_j\ra_NC\big)+\tau(A_j-A)\tau(B^{\dag}C),\cr}
$$
and choose $j$ such that $\Vert A-A_j\Vert\leq\epsilon/(3\Vert
B\Vert\Vert C\Vert)$. Now choose $N_0=N_0(j)$ such that
$|\tau\big(B^{\dag}\la A_j\ra_NC-\tau(A_j)B^{\dag}C\big)|<\epsilon/3$,
for all $N>N_0$. Then, using \timeavcont ,
$$
\eqalign{
|\tau\big(B^{\dag}\la A\ra_NC-\tau(A)B^{\dag}C\big)|&<
\epsilon/3+\Vert\la A-A_j\ra_N\Vert\Vert B\Vert\Vert C\Vert+\Vert A_j-A\Vert
\Vert B\Vert\Vert C\Vert\cr
&<2\Vert A-A_j\Vert\Vert B\Vert+\epsilon/3\leq\epsilon,\cr}
$$
for all $N>N_0$. We have thus shown that \wconv\ holds for all
$A\in\fA$ and $B,C\in\fA_{loc}$. Repeating twice almost {\it verbatim}
the above $3\epsilon$ argument we establish \wconv\ for all
$A,B,C\in\fA$.

\noindent
$2^{\circ}$. Having established \timeavconvone\ for $\phi=[B],
\;\psi=[C]$, and $A\in\fA$, we now show that it holds for
arbitrary $\phi,\;\psi\in\fH$. Let $\phi\in\fH$, and let $B_j$
be such that $\Vert\phi-[B_j]\Vert<\epsilon/(2\Vert[C]\Vert
\Vert A\Vert)$. Write
$$
\big(\phi,(\la A\ra_N-\tau(A)I)[C]\big)=
\big([B_j],(\la A\ra_N-\tau(A)I)[C]\big)+
\big(\phi-[B_j],(\la A\ra_N-\tau(A)I)[C]\big),
$$
and choose $N_0=N_0(j)$ so that $|\big([B_j],(\la A\ra_N-\tau(A)I)
[C]\big)|<\epsilon/2$, for all $N>N_0$. Then
$$
|\big(\phi,(\la A\ra_N-\tau(A)I)[C]\big)|<\epsilon/2+
\Vert\phi-[B_j]\Vert\Vert[C]\Vert\Vert A\Vert<\epsilon.
$$
Repeating this argument we establish \timeavconvone\ for all
$\psi\in\fH$. The proof of the proposition is complete.

\section\quantentsec{Entropy of a quantum source}

\newsubsec
In this subsection we construct the entropy of a quantum source.
Our construction is largely standard, see e.g. [LR] and [W], but
we include most of the details to make the presentation self contained.
The key mathematical input is the following lemma whose proof can
be found on page 1122 of [LR].

\lemma\kleinlemma{Let $A$ and $B$ be positive trace class operators
on a Hilbert space. Then
$$
\tr(A\log A-A\log B)\geq\tr(A-B).\ref{\kleinineq}
$$}

We define the entropy associated with a sequence of $n$ signals
to be
$$
H_n(\Pi):=-\tr_{\cH^{\otimes n}}(\Pi_n\log\Pi_n).\num
$$
Substituting $A=\Pi_{m+n}$ and $B=\Pi_m\otimes\Pi_n$
in \kleinineq , we obtain the following subadditivity property
of $H_n(\Pi)$:
$$
H_{m+n}(\Pi)\leq H_m(\Pi)+H_n(\Pi).\ref{\subadditivity}
$$
A standard argument, see e.g. [K], pages 48--49, shows that the
limit
$$
h(\Pi)=\lim_{n\to\infty}\;{1\over n}\;H_n(\Pi)\ref{\entropydef}
$$
exists. We call $h(\Pi)$ the entropy of the quantum source. Notice
also that subadditivity implies the inequality
$$
h(\Pi) \leq -\tr_{\cH}(\rho\log\rho) \leq \log d,\ref{\entropybound}
$$
where $\rho$ is the density matrix for the signals, and $d$ is the
dimension of the signal Hilbert space $\cH$.

\subsec
We can easily compute the entropy of two basic types of quantum
sources introduced in {\quantsourcesec}.

For a Bernoulli source, $\Pi_n=\rho\otimes\ldots\otimes\rho$, and so
$$
\eqalign{
{1\over n}\;H_n(\Pi)
&=-{1\over n}\;\tr_{\cH^{\otimes n}}\big((\otimes^n\rho)\;(\log
\otimes^n\rho)\big)\cr
&=-{1\over n}\;\sum_{1\leq j\leq n}\tr_{\cH^{\otimes n}}\big((
\otimes^n\rho)\;(I\otimes\ldots\otimes\log\rho\otimes\ldots\otimes
I)\big)\cr
&= -\tr_{\cH}(\rho\log\rho).\cr}
$$
As a result, $h(\Pi)=-\tr(\rho\log\rho)$, i.e. the entropy of a
Bernoulli source is equal to the von Neumann entropy of the signal
density matrix.

For the non-Bernoulli source defined in   {\commutingR} a similar
calculation yields
$h(\Pi) = -\tr((\rho \otimes I) R \log R)$. We have not found a
closed form expression for
the entropy of the other non-Bernoulli source described
in {\quantsourcesec}.

\section\equipartsecsec{Asymptotic equipartition property}

\newsubsec
The classical asymptotic equipartition property, also known as the
Shannon-McMillan theorem, see e.g. [A], [K], states that if $\csrc$
is an ergodic source with entropy $h(\mu)$, then the sequence
$$
f_n(\ul x):=-{1\over n}\log\mu(\{\ul y\in\cX^\infty:\; y_1=x_1,\ldots,
y_n=x_n\})\ref{\empentref}
$$
converges in measure to $h(\mu)$. In other words, given $\delta,
\epsilon>0$, there is $n_0$ such that for all $n\geq n_0$,
$$
\mu(\{\ul x\in\cX:\; |f_n(\ul x)-h(\mu)|>\delta\})<\epsilon.
\ref{\mcmillan}
$$
This is interpreted as saying that for any length $n$,
there are two categories of
messages sent by a source: ({\it i}) a small fraction of ``likely''
messages, each of which carries equal probability
$$
\mu(\{\ul y\in\cX^\infty:\; y_1=x_1,\ldots,y_n=x_n\})\sim
e^{-nh(\mu)},\num
$$
and ({\it ii}) the bulk of ``unlikely'' messages whose total
probability goes to zero as $n$ goes to infinity. There are approximately
$e^{nh(\mu)}$ likely messages which is much less than the total
number of messages $e^{n\log r}$ (unless $h(\mu)$ happens to
equal $\log r$).

The goal of this section is to establish an analogous result for
quantum sources. Our theorem generalizes Schumacher's result [S],
[JS] to general, not necessary Bernoulli, sources.

\subsec
Let $\bfA=\{A_1,\ldots,A_r\}$, $r<\infty$ be a family of observables
on $\cH$ such that $A_j\geq 0$, for all $j$, and
$$
A_1+\ldots+A_r=I.\ref{\sumofobs}
$$
Such a family is called a {\it positive operator valued measure}
(POM), see [HJW] and references therein. A POM is called {\it pure}
if each $A_J$ is a rank one operator. For example, any family of
$d=\dim\cH$ pairwise orthogonal projections on $\cH$ satisfies the
above conditions and so is a pure POM. We will call
the set $\cX_{\bfA}=\{1,\ldots,r\}$ the classical alphabet associated
with the POM $\bfA$, and denote by $\cX_{\bfA}^\infty$
the space of all infinite messages over the alphabet $\cX_{\bfA}$.
We can define a probability measure on $\cX_{\bfA}^\infty$ associated
with the quantum source $(\fA,\tau,\alpha)$. For $\{k_1,\ldots,k_n\}
\in\cX^n$ we define
$$
\mu^{\bfA}_n(\{\ul x:\; x_1=k_1,\ldots,x_n=k_n\}):=\tau_n(A_{k_1}
\otimes\ldots\otimes A_{k_n}).\ref{\mundef}
$$
This defines a consistent family of cylinder measures, as
$$
\eqalign{
\sum_{k\in\cX_{\bfA}}\mu^{\bfA}_{n+1}(\{\ul x:\; x_1=k_1,\ldots,\;
&x_n=k_n,\;x_{n+1}=k_{n+1}\})\cr
&=\tau_{n+1}(A_{k_1}\otimes\ldots\otimes
A_{k_n}\otimes\sum_{k\in\cX_{\bfA}}A_k)\cr
&=\tau_{n+1}(A_{k_1}\otimes\ldots\otimes A_{k_n}\otimes I)
=\tau_n(A_{k_1}\otimes\ldots\otimes A_{k_n})\cr
&=\mu^{\bfA}_n(\{\ul x:\; x_1=k_1,\ldots,x_n=k_n\}).\cr}
$$
Let $\mu^{\bfA}$ denote the probability measure obtained from
$\{\mu^{\bfA}_n\}$ by means of Kolmogorov's extension theorem. By
$T$ we denote the shift operator on $\cX_{\bfA}^\infty$. Then the
triple $(\cX_{\bfA}^\infty,T,\mu^{\bfA})$ forms a classical
information source. We emphasize that it depends on the choice of
$\bfA$. Obviously, this source is stationary.

\subsec
In fact, this source is ergodic if the underlying quantum source is
ergodic. We state it as the following lemma.
\lemma\classerg{If $\qsrc$ is ergodic, then for any choice of $\bfA$, the
classical source constructed above is ergodic.}
\proof We show that $(\cX_{\bfA}^\infty,T,\mu^{\bfA})$ has the following
property: for every $f\in L^1(\cX^\infty,d\mu^{\bfA})$, the sequence
$\la f\ra_N$ converges in $L^1$ to $\int f\;d\mu^{\bfA}$. Using Fatou's
lemma, and recalling the individual ergodic theorem, this
implies that the only functions invariant under $T$ are constants,
and so the classical source is ergodic. We first observe that
it is sufficient to show that if $\cC$ is a cylinder set, and
$\chi_{\cC}$ denotes the corresponding characteristic function, then
$$
\int_{\cX_{\bfA}^\infty}|\la\chi_{\cC}\ra_N(\ul x) -
\mu^{\bfA}(\cC)| d\mu^{\bfA}(\ul x)\longrightarrow0,
\ref{\convergeinL1}
$$
as $N\to\infty$. A standard ``${\epsilon / 3}$" argument then implies
that for all $f\in L^1(\cX^\infty,d\mu^{\bfA})$, $\la f\ra_N$
converges to $\int f\;d\mu^{\bfA}$ in the $L^1$-norm. Furthermore,
since $d\mu^{\bfA}$ is a probability measure, {\convergeinL1} will
follow from convergence of $\la\chi_{\cC}\ra_N$ to $\mu^{\bfA}(\cC)$ in
the $L^2$-norm. This in turn is implied by the following stronger
result. Suppose $\cC$ and $\cD$ are cylinder
sets, and $\chi_{\cC}$ and $\chi_{\cD}$ are the corresponding
characteristic functions; then
$$
\int_{\cX_{\bfA}^\infty}\la\chi_{\cC}\ra_N(\ul x)\chi_{\cD}(\ul x)\;
d\mu^{\bfA}(\ul x)\longrightarrow\mu^{\bfA}(\cC)\mu^{\bfA}(\cD),
\ref{\toshow}
$$
as $N\to\infty$.

In order to prove {\toshow}, let
$\cC=\{\ul x:\;x_1=k_1,\ldots,\;x_n=k_n\}$, $\cD=\{\ul x:\;
x_{j+1}=l_1,\ldots,\;x_{j+m}=l_m\}$. The corresponding observables
are given by
$$
G(\cC) = A_{k_1}\otimes\ldots\otimes A_{k_n},\quad\quad
G(\cD) = A_{l_1}\otimes\ldots\otimes A_{l_m}.
$$
There is $N_0$ such that
for all $n>N_0$, $T^n(\cC)\cap\cD=\emptyset$. Therefore,
$$
\eqalign{
{1\over N}\int_{\cX_{\bfA}^\infty}\;\sum_{i=N_0+1}^{N-1}
&\chi_\cC(T^i\ul x)\chi_\cD(\ul x)\;d\mu^{\bfA}(\ul x)\cr
&={1\over N}\;\tau\big(\sum_{i=N_0+1}^{N-1}\alpha^i
(G(\cC)) G(\cD)\big)\cr
&=\tau\big(\langle G(\cC)\rangle_{N}) G(\cD)\big)
- {1\over N}\;\tau\big(\sum_{i=0}^{N_{0}}\alpha^i
(G(\cC)) G(\cD)\big).\cr}
$$
By our assumption of quantum ergodicity, the first term above
converges to $\mu^{\bfA}(\cC)\mu^{\bfA}(\cD)$,
as $N\rightarrow\infty$. The second term is bounded as follows:
$$
\Bigl|{1\over N}\;\tau\big(\sum_{i=0}^{N_{0}}\alpha^i
(G(\cC)) G(\cD)\big)\Bigr| \leq
{N_{0} + 1\over N} \prod_{p=1}^{n} \Vert A_{k_{p}} \Vert\,
\prod_{q=1}^{l} \Vert A_{l_{q}} \Vert,
$$
which converges to 0 as $N\rightarrow\infty$. Finally,
$$
{1\over N}\int_{\cX_{\bfA}^\infty}\;\sum_{i=0}^{N_0}
\chi_\cC(T^i\ul x)\chi_\cD(\ul x)\;d\mu^{\bfA}(\ul x)\leq
{N_{0} + 1\over N}
\longrightarrow 0,
$$
and the proof is complete.

\subsec
Let $h_{\bfA}$ denote the Shannon entropy of the classical
source $(\cX_{\bfA}^\infty,T,\mu^{\bfA})$ ([B], [K]). In other
words,
$$
\eqalign{
h_{\bfA}&=-\lim_{n\to\infty}{1\over n}\sum_{k_1,\ldots,k_n\in
\cX_{\bfA}}\mu_{\bfA}(\{\ul x:\;x_1=k_1,\ldots,x_n=k_n\})\cr
&\hskip 3.7cm\times\log\mu_{\bfA}(\{\ul x:\;x_1=k_1,\ldots,x_n=k_n\})\cr
&=-\lim_{n\to\infty}{1\over n}\sum_{k_1,\ldots,k_n\in
\cX_{\bfA}}\tau(A_{k_1}\otimes\ldots\otimes A_{k_n})
\log\tau(A_{k_1}\otimes\ldots\otimes A_{k_n}).\cr}
$$
A remarkable fact about $h_{\bfA}$ is that it can be bounded
from below in terms of $h(\Pi)$, the quantum entropy of the source,
and a quantity depending exclusively on the statistical properties
of the signal ensemble.

\thm\quantbound{For any POM $\bfA$ the following inequality holds:
$$
h_{\bfA}\geq h(\Pi)-\sum_{1\leq k\leq r}\tr(A_k\rho)\log\tr(A_j).
\ref{\inequality}
$$}

\noindent
{\bf Remark}. In particular, if $\bfA$ is a pure POM consisting of $d$
mutually orthogonal projections, then $h_{\bfA}\geq h(\Pi)$.
\proof The proof of this theorem is based on the following lemma.
\lemma\basicineqlemma{Let $\cF$ be a Hilbert space of finite dimension
$N$, and let $R$ be a density matrix on $\cF$. If $A_1,\ldots,\;A_r$
are positive operators such that $A_1+\ldots+A_r=I$, then
$$
\sum_{1\leq j\leq r}\tr(A_jR)\log\tr(A_jR)\leq
\tr(R\log R)+\sum_{1\leq j\leq r}\tr(A_jR)\log\tr(A_j).
\ref{\basicinequality}
$$}
\proof Let $A_j^{kl}$ denote the matrix entries of $A_j$ in an
orthonormal basis consisting of eigenvectors of $R$. Then
$\tr(A_jR)=\sum_k\lambda_kA^{kk}_j$, where $\lambda_1,\ldots,
\lambda_N$ are the eigenvalues of $R$. The function $[0,1]\ni
x\rightarrow f(x):=x\log x$ is convex, and so by Jensen's inequality,
$$
\eqalign{
\sum_jf\big(\sum_k\lambda_kA^{kk}_j\big)
&=\sum_j\tr(A_j)f\big(\sum_k{{A^{kk}_j}\over{\tr(A_j)}}
\lambda_k\big)+\sum_j\sum_k{{A^{kk}_j}\over{\tr(A_j)}}
\lambda_k\;f(\tr(A_j))\cr
&\leq\sum_j\sum_kA_j^{kk}f(\lambda_k)+\sum_k\tr(A_jR)\log\tr(A_j)\cr
&=\sum_kf(\lambda_k)+\sum_k\tr(A_jR)\log\tr(A_j),\cr}
$$
which implies {\basicinequality}.

As a consequence of this lemma, and condition $3^\circ$ in the
definition of a consistent family of density matrices,
$$
\eqalign{
&\sum_{k_1,\ldots,k_n\in\cX_{\bfA}}\tr\big((A_{k_1}\otimes\ldots
\otimes A_{k_n})\Pi_n\big)\log\tr\big((A_{k_1}\otimes\ldots\otimes
A_{k_n}\Pi_n)\big)\cr
&\qquad\leq\tr(\Pi_n\log\Pi_n)+
\sum_{k_1,\ldots,k_n\in\cX_{\bfA}}\tr\big((A_{k_1}\otimes\ldots
\otimes A_{k_n})\Pi_n\big)\log\tr(A_{k_1}\otimes\ldots\otimes
A_{k_n})\cr
&\qquad=\tr(\Pi_n\log\Pi_n)+\sum_{1\leq j\leq n}\;
\sum_{k_1,\ldots,k_n\in\cX_{\bfA}}\tr\big((A_{k_1}\otimes\ldots
\otimes A_{k_n})\Pi_n\big)\log\tr(A_{k_j})\cr
&\qquad=\tr(\Pi_n\log\Pi_n)+n\sum_{1\leq k\leq r}
\tr(A_k\rho)\log\tr(A_k),\cr}
$$
and the claim follows.

\subsec
The theorem below is the main result of this section. It can be
regarded as a quantum version of the Shannon-McMillan theorem.

\thm\shanmcmillthm{Let $(\fA,\tau,\alpha)$ be an ergodic source,
and let $\bfA$ be a POM, for which the operators
$\{A_{j}\},\; 1 \leq j \leq r$, are orthogonal projections.
Let $h_\bfA$ be the Shannon entropy
of the associated classical source, and define
$$
M = \max_{1 \leq j \leq r} {\rm rank} (A_{j}), \quad \quad
m = \min_{1 \leq j \leq r} {\rm rank} (A_{j}).
$$
Then, given $\delta,\epsilon>0$,
there is $n_0$ such that for all $n\geq n_0$,
$$
\cH^{\otimes n}=\cS_n\oplus\cS_n^\perp,
$$
where $\cS_n$ is a subspace whose dimension satisfies
$$
{{\log m} - \delta \over {\log d}} \leq
{{\log\dim\cS_n}\over{\log\dim\cH^{\otimes n}}} -
{h_{\bfA}\over {\log d}} \leq
{{\log M} + \delta \over {\log d}}.\ref{\deltaineq}
$$
Further, let $P_{\cS_n}$ be the orthogonal projection onto $\cS_n$. Then
for any observable $C\in\cL(\cH^{\otimes n})$,
$$
|\tau(CP_{\cS_n})-\tau(C)|<\epsilon\Vert C\Vert.\ref{\epsilonineq}
$$}

\smallskip\noindent
{\bf Remark 1.} We can regard $\cS_n$ as a significant subspace of
$\cH^{\otimes n}$ in the sense that the expectation of any observable
is almost completely determined by its restriction to $\cS_n$.
If the states $\{|\psi_{j}\rangle\}$ are orthogonal, we can take
the $d$ orthogonal projections $\{|\psi_{j}\rangle \langle\psi_{j}|\}$
for the POM. In this case the entropy $h_{\bf A}$ is equal to the
von Neumann entropy, since the density operators all commute.
Then there is a direct correspondence with the classical Shannon-McMillan
theorem, and the quantum theory is just a restatement of the
classical result.

\smallskip\noindent
{\bf Remark 2.} In the Bernoulli case we can take the POM $\bfA$ to be
$d$ orthogonal projections onto the eigenvectors of $\rho$,
in which case the
inequality of {\quantbound} is saturated. If $\rho$ has
simple spectrum, this means that the Shannon entropy equals
the von Neumann entropy of the quantum source. Our
result then agrees with Schumacher's conclusion that the
information contained in the quantum source resides in a
subspace whose dimension is asymptotically
$e^{n h_{\bfA}}$. In the general case we obtain
only an upper bound for the dimension of the relevant subspace, and
this upper bound depends on the choice of POM. For example,
if each operator $A_{j}$ in the POM $\bfA$ is equal to
$(1/d)I$ where $I$ is the identity, then the Shannon entropy is
$h_{\bfA} = \log d$. Since this is the maximum possible entropy
for a POM with $d$ operators, we can conclude that all
information about the quantum source has been lost in this
measurement process. As these results
show, it is advantageous to use a POM composed of orthogonal
projections.

\proof
We use the POM to construct the classical source
$\cX^{\infty}_{\bfA}$, with entropy $h_{\bfA}$.
Let $f_{n}({\ul x})$ be the empirical
entropy of a message $\ul x$ of length $n$ defined in
{\empentref}. Given $\epsilon, \delta > 0$ let us define the sets
$$
U_{n, \delta} = \{{\ul x} \in \cX \,:\,
|f_{n}({\ul x}) - h_{\bfA}| > \delta \},
$$
$$
L_{n, \delta} = \{{\ul x} \in \cX \,:\,
|f_{n}({\ul x}) - h_{\bfA}| \leq \delta \}.
$$
By the Shannon-McMillan theorem, given
$\epsilon, \delta >0$, there is $n_{0}$ such that for all
$n \geq n_{0}$,
$$
\sum_{{\ul x} \in U_{n, \delta}} {\mu}_{\bfA}(\{{\ul x}\})
< \epsilon.
$$
Since the operators $\{A_{j}\}$ are orthogonal projections,
each tensor product $A_{k_{1}} \otimes \dots \otimes A_{k_{n}}$
is an orthogonal projection, and hence so is the sum of
these operators over the set $L_{n, \delta}$. Let
$\cS_n$ denote the range of this projection, and let
$P_{\cS_n}$ denote the orthogonal projection onto this
subspace. Then for any observable $C$, we have
$$
|\tau(C P_{\cS_n}) - \tau(C)| \leq \epsilon \Vert C\Vert.
$$

It remains to estimate the dimension of $\cS_n$. Since the
projections $\{A_{j}\}$ are orthogonal, its dimension is
given by
$$
{\rm dim} (\cS_n) = \sum_{{\bf x} \in L_{n, \delta}}
\prod_{j=1}^{n} {\rm rank} (A_{x_{j}}).
$$
Therefore,
$$
m^{n} \bigl|L_{n, \delta}\bigr| \leq
{\rm dim} (\cS_n) \leq M^{n} \bigl|L_{n, \delta}\bigr|,
$$
where $\bigl|L_{n, \delta}\bigr|$ is the size of the set
$L_{n, \delta}$. The Shannon-McMillan theorem implies that
$$
(1 - \epsilon) e^{n(h_{\bf A} + \delta)} \geq
\bigl|L_{n, \delta}\bigr|\geq
e^{n(h_{\bf A} - \delta)}.
$$
This leads to
$$
{{\log m} \over {\log d}} - {\delta \over {\log d}} \leq
{{\rm dim} (\cS_n) \over n {\log d}} -
{h_{\bf A} \over {\log d}} \leq
{{\log M} \over {\log d}} + {\delta \over {\log d}}
+ {{\log (1-\epsilon)} \over n},
$$
and the result follows.

\appendix\markovapp
\medskip
We present here the proof that the $R$-matrix {\twobytwoR} defines
a family of positive density matrices, when $|b|$ and $|c|$ are
sufficiently small. It is convenient to introduce the matrices
$$
\omega = {a \over 2}I - {1 \over 2}{\sigma}_{3},
$$
$$
Q =  {\rho}^{-1} (b{\sigma}_{1} + c{\sigma}_{2}).
$$
Note that we assume $|a|<1$, so ${\rho}^{-1}$ exists. Then for all
$n \geq 2$ we define
$$
S_{n} = I_{n-2} \otimes \omega \otimes Q.
$$
It follows by direct calculation that for all $n \geq 1$,
$$
{\Pi}_{n+1} = {\Pi}_{n} \otimes \rho +
{1 \over 2} ({\Pi}_{n} \otimes \rho) S_{n+1} +
{1 \over 2} S_{n+1}^{\dagger} ({\Pi}_{n} \otimes \rho).
$$
In order to proceed we
make the inductive assumption that ${\Pi}_{n} > 0$;
this implies in particular that $({\Pi}_{n} + u)^{-1}$
is bounded for every $u \geq 0$. We will prove that
$({\Pi}_{n+1} + u)^{-1}$ is bounded for every $u \geq 0$; together
with the positivity of ${\Pi}_{1} = \rho$,
this will establish the desired result.

Note first that $\Vert S_{n}\Vert \leq \Vert\omega\Vert\,\Vert Q\Vert$,
and this bound is uniform in $n$. For convenience we define
$$
w = \Vert \omega\Vert ,\quad\quad q = \Vert Q\Vert .
$$
Furthermore, for any $u \geq 0$,
$$
\eqalign{
({\Pi}_{n+1} + u)^{-1} =  &({\Pi}_{n} \otimes \rho + u)^{-1}
(I + {1 \over 2} ({\Pi}_{n} \otimes \rho) S_{n+1}
({\Pi}_{n} \otimes \rho + u)^{-1} \cr
&+{1 \over 2} S_{n+1}^{\dagger} ({\Pi}_{n} \otimes \rho)
({\Pi}_{n} \otimes \rho + u)^{-1})^{-1}.\cr}
$$
Our inductive assumption implies that $({\Pi}_{n} \otimes \rho + u)^{-1}$
is bounded for $u \geq 0$. Choosing $|b|, |c|$ sufficiently
small guarantees that  $\Vert S_{n}^{\dagger}\Vert  < \epsilon$ for any
$\epsilon > 0$. Furthermore,
$$
({\Pi}_{n} \otimes \rho) S_{n+1} ({\Pi}_{n} \otimes \rho + u)^{-1}
= \Bigl([{\Pi}_{n} (I_{n-1} \otimes \omega) {\Pi}_{n}^{-1}]
\otimes Q^{\dagger}\Bigr) ({\Pi}_{n} \otimes \rho)
({\Pi}_{n} \otimes \rho + u)^{-1}.
$$
Since $\Vert Q^{\dagger}\Vert = q$ can be made arbitrarily small by choosing
$|b|, |c|$ sufficiently small, the boundedness of
$({\Pi}_{n+1} + u)^{-1}$ will follow from a bound for the
operator ${\Pi}_{n} (I_{n-1} \otimes \omega) {\Pi}_{n}^{-1}$ which
is uniform in $n$. Accordingly let us define for $n \geq 1$,
$$
A_{n} = {\Pi}_{n} (I_{n-1} \otimes \omega) {\Pi}_{n}^{-1}.
$$
By imitating the derivation above, we obtain the recursion relation
$$
A_{n} = \Bigl(I + {1 \over 2} A_{n-1} \otimes Q^{\dagger} +
{1 \over 2} S_{n}^{\dagger}\Bigr) (I_{n-1} \otimes \omega)
\Bigl(I + {1 \over 2} A_{n-1} \otimes Q^{\dagger} +
{1 \over 2} S_{n}^{\dagger}\Bigr)^{-1}.\ref{\recursionforA}
$$

It is immediate that $\Vert A_{1}\Vert  =
\Vert \omega\Vert  \leq 1$. We make the
inductive assumption that $\Vert A_{n-1}\Vert  \leq 1$; then
{\recursionforA} implies the estimate
$$
\Vert A_{n}\Vert  \leq (1 + {q \over 2} + {w q \over 2}) w
(1 - {q \over 2} - {w q \over 2})^{-1}.\ref{\boundforA}
$$
If we choose
$$
q < 2 {1 - w \over (1 + w)^{2}},\ref{\boundforq}
$$
then {\boundforA} implies that $\Vert A_{n}\Vert  \leq 1$.
Hence by choosing {\boundforq} we obtain that $A_{n}$ is uniformly
bounded for all $n$, and hence that $({\Pi}_{n+1} + u)^{-1}$
is bounded for all $u \geq 0$. Therefore the positivity of the
density matrices is proved.

\medskip\noindent
{\bf Acknowledgement.} We would like to thank Evelyn Wright for
helpful discussions.
\vfill\eject

\centerline{\bf References}
\baselineskip=12pt
\frenchspacing

\bigskip
\item{[A]} Ash, R. B.: {\it Information Theory}, Dover (1990)
\item{[Be]} Bennett, C. H.: Quantum Information and
Computation, {\it Physics Today}, October 1995, 24--30
\item{[B]} Billingsley, P.: {\it Ergodic Theory and Information},
Robert E. Krieger Publishing Company (1965)
\item{[BR]} Bratteli, O., and Robinson, D. W.: {\it Operator
Algebras and Quantum Statistical Mechanics}, vols 1 $\&$ 2, Springer
Verlag (1981)
\item{[CFS]} Cornfeld, I. P., Fomin, S. V., and Sinai, Ya.:
{\it Ergodic Theory}, Springer Verlag (1982)
\item{[H]} Haag, R.: {\it Local Quantum Physics}, Springer
Verlag (1992)
\item{[HJW]} Hughston, L. P., Jozsa, R., and Wootters, W. .:
A complete classification of quantum ensembles having a given
density matrix, {\it Phys. Lett.}, {\bf A183}, 14--18 (1993)
\item{[JS]} Jozsa, R., and Schumacher, B.: A new proof of the
quantum noiseless coding theorem, {\it J. Mod. Optics},
{\bf 41}, 2343--2349 (1994)
\item{[K]} Khinchin, A. I.: {\it Mathematical Foundations of
Information Theory}, Dover (1957)
\item{[LR]} Lanford, O. E., and Robinson, D. W.: Mean entropy
of states in quantum-statistical systems, {\it J. Math. Phys.},
{\bf 9}, 1120--1125 (1968)
\item{[S]} Schumacher, B.: Quantum coding, {\it Phys. Rev. A},
{\bf 51}, 2738--2747 (1995)
\item{[W]} Wehrl, A.: General properties of entropy, {\it Rev.
Mod. Phys.}, {\bf 50}, 221--260 (1978)

\vfill\eject\end